\documentstyle[prl,aps,graphicx,multicol]{revtex}
\input psfig.sty
\begin{document}

\title{Fermi surface and heavy masses for UPd$_2$Al$_3$}

\author{G. Zwicknagl$^1$, A. Yaresko$^2$ and P. Fulde$^3$}

\address{
1 Institut f\"ur Mathematische Physik, Technische Universit\"at 
Braunschweig\\ Mendelssohnstr. 3, 38106 Braunschweig, Germany\\
2 Max-Planck-Institut f\"ur Chemische Physik fester Stoffe\\
N\"othnitzer Stra\ss e 40, 01187 Dresden, Germany\\
3 Max-Planck-Institut f\"ur Physik komplexer Systeme\\
N\"othnitzer Stra\ss e 38, 01187 Dresden, Germany}

\date{\today}
\maketitle

\begin{abstract}
We calculate the Fermi surface and the anisotropic heavy masses of
UPd$_2$Al$_3$ by keeping two of the 5$f$ electrons as localized. Good agreement
with experiments is found. The theory contains essentially no adjustable
parameter except for a small shift of the position of the Fermi energy of the
order of a few meV. A discussion is given why localization of two $f$ electrons
is justified. 
\end{abstract}

%\noindent PACS.~~~~~~~~
%\begin{minipage}[t]{15cm}
%\end{minipage}

\vspace{1.5cm}

\newcommand{\gsim}{\mathrel{\raise.3ex\hbox{$>$\kern-.75em\lower1ex\hbox{$\sim$}}}}
\newcommand{\lsim}{\mathrel{\raise.3ex\hbox{$<$\kern-.75em\lower1ex\hbox{$\sim$}}}}

\begin{multicols}{2}
\narrowtext
%\section{Introduction}

There has been experimental evidence since a number of years that in
UPd$_2$Al$_3$ the 5$f$ electrons have a dual character. The large observed
magnetic moment of 0.85 $\mu_B$ \cite{KFRMGSGLS} suggests localization of the
$f$ electrons. On the other hand, the large jump in the specific heat at the
superconducting transition temperature T$_c$=2K implies a large Sommerfeld
specific heat coefficient $\gamma$ and moderately heavy fermion behavior
\cite{CHKSWKGSS}. This requires delocalized 5$f$ electrons. The dual character
is also found in a number of other physical properties such as photoemission,
inelastic neutron scattering and muon spin rotation
\cite{TSYCMK,MeHaKoOn,BSRAHLEK,FAGSGSSK}. It is also observed in other metallic
U compounds \cite{ScVoLoHuMa,YDRGKMMC}. Moreover, the assumption is supported
by quantum-chemical calculations on U(C$_8$H$_8$)$_2$ \cite{lanthanocenes}
which exhibit a number of low-lying excitations caused by intra-atomic
rearrangements of the 5$f$ electrons. There are speculations that similar 5$f$
excitations might even be responsible for the formation of Cooper pairs in
superconductors \cite{UPd2Al3Nature}. In that case the dual model should allow
for a natural description of heavy-fermion superconductivity coexisting with
magnetism produced by 5$f$ electrons.

We have previously applied the dual model in order to explain the mass
enhancement of the quasiparticle excitations and the de Haas-van Alphen
frequencies in UPt$_3$, the model system of heavy-fermion behavior in U
compounds \cite{ZwYaFu}. Our theory conjectured that the delocalized 5$f$
states hybridize with the conduction electron states and form energy bands
while the localized ones form multiplets in order to reduce the local mutual
Coulomb repulsions. The interaction of the two subsystems, i.e., the
delocalized with the localized one leads to a mass enhancement of the
former. The situation resembles that of Pr metal where a mass enhancement of
the conduction electrons by a factor of 5 results from virtual crystal-field
excitations of the localized 4$f$ electrons \cite{WhiteFulde}. The same
mechanism leads to the heavy quasiparticles in the recently discovered
heavy-fermion superconductor PrOs$_4$Sb$_{12}$ \cite{BFHZM}. The role of the
incompletely filled 4$f$ shell is taken by an incompletely filled subshell of
5$f$ electrons when U ions are considered instead.

The coexistence of itinerant and localized 5$f$ states is referred to as
partial localization. It plays an important role in many intermetallic actinide
compounds. Partial localization arises from an interplay between the
hybridization of the 5$f$ states with the environment and on-site Coulomb
correlations. This is discussed below in more detail. The underlying
microscopic origin is an area of active current research
\cite{EfHaRuFuZw,LuSaErJo,SoAhErJoWi}. The associated physical picture is
therefore quite different from the one suggested in Ref. \cite{CHKSWKGSS},
where two disjunct subsystems in $\bf k$-space were postulated.

The dual model should be contrasted with density functional based calculations
which use the local density approximation. These have
been successful in explaining the measured de Haas-van Alphen frequencies of
systems like UPd$_2$Al$_3$ or UPt$_3$ \cite{KnMaSaKu,KimuraUPt33,IYHSTHYOY}, but they fail to
explain the observed heavy quasiparticle masses. For  UPt$_3$ the observed
masses are by a factor of 20 larger than the calculated ones and for
UPd$_2$Al$_3$ the difference is roughly a factor of 4. When
using the self-interaction corrected local spin-density approximation
(SIC-LSDA) \cite{PeTeSzTy} a ground state is found with coexisting
localized (f$^2$) and delocalized 5$f$ electrons of U. But the calculated
density of states is too small by a factor of approximately 10 to
account for the observed linear specific heat.

While the calculated energy bands are too broad for explaining the effective
masses, they are too small in order to fit the observed photoemission data of
UPt$_3$, UPd$_2$Al$_3$ or UBe$_{13}$ \cite{JWAllenPESReview}. The latter shows
a broad peak just below the Fermi energy E$_{\rm F}$ and is quite different
from the data of heavy quasiparticle systems which involve Ce$^{3+}$ instead of
U ions. Here a broad peak is found approximately 2eV below E$_{\rm F}$
\cite{KondoRes} while at the Fermi energy a small Kondo resonance is
detected. We add that even for UPd$_3$ which has localized 5$f$ electrons and
does not show heavy-fermion behavior, the photoemission data resemble that of
the previously mentioned U compounds, except that there is no $f$ weight at
E$_{\rm F}$. This indicates that despite the afore mentioned successes the
strong correlations in those materials are not properly treated by the
presently used computational methods. Instead we find that a more microscopic
understanding of heavy-fermion behavior is highly desirable.

The aim of the present investigation is to show that the de Haas-van Alphen
frequencies for the heavy-quasiparticle portion of the Fermi surface and
the large effective masses (including their anisotropies), can be explained
very well by treating two of the 5$f$ electrons as being localized. We put the
localized electrons into 5$f$ j=$\frac{5}{2}$ orbitals with j$_z=\pm
\frac{5}{2}$ and $\pm \frac{1}{2}$. The j$_z=\pm \frac{3}{2}$ states are
treated as itinerant electrons. The reason for this choice is explained below,
but it is worth pointing out here that the j$_z=\pm \frac{3}{2}$ states
hybridize strongest among the different ones in a conventional LDA calculation.  

\begin{figure}
\vspace{-0.5cm}
\hspace{1cm}
\includegraphics[width=6.2cm]{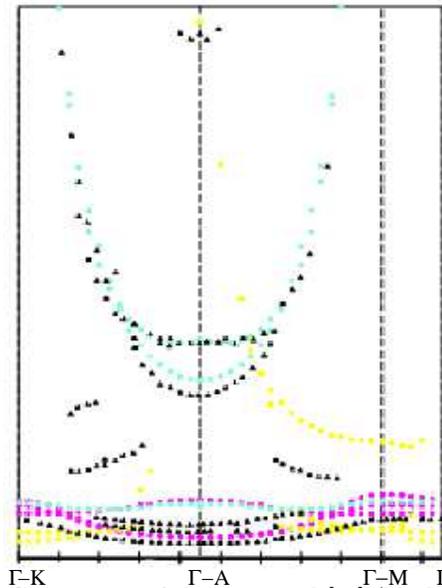}
%\vspace{-0.5cm}
\caption{Comparison of the measured [19] (black symbols) and
calculated de Haas-van
Alphen frequencies for UPd$_2$Al$_3$. The dominant part reflects a nearly
cylindrical sheet of the Fermi surface.}
\label{fig:Fig1}
\end{figure}

The calculations of the heavy bands proceed in three steps. First, the
bandstructure is determined by solving the Dirac equation for the
self-consistent LDA potentials, thereby excluding the U 5$f$ states with
j=$\frac{5}{2}$, j$_z=\pm \frac{5}{2}$ and j$_z=\pm \frac{1}{2}$ from forming
bands. Two 5$f$ electrons in localized orbitals are accounted for in the
self-consistent density and, concomitantly, in the potential seen by the
conduction electrons. The intrinsic bandwidth of the itinerant U 5$f$
j=$\frac{5}{2}$, j$_z=\pm \frac{3}{2}$ electrons is taken from the LDA
calculation while the position of the corresponding band center C is chosen
such that the density distribution of the conduction electrons remains the same
as within the LDA. The $f$ electron count per U atom for the delocalized 5$f$
electrons amount to n$_{\rm f} \simeq \frac{2}{3}$, i.e., the system is of a
mixed valence type. The calculated de Haas-van Alphen frequencies are shown in
Fig. \ref{fig:Fig1} which also contains the experimental results. 

\vspace{1.8cm}
%%%%%%%%%%%%%%%%%%%%%%%%%%%%%%%%%%%%%%%%%%%%%%%%%%%%%%%%%%%%%%%%%%%%%%%%%%%%
%Figur 1
%%%%%%%%%%%%%%%%%%%%%%%%%%%%%%%%%%%%%%%%%%%%%%%%%%%%%%%%%%%%%%%%%%%%%%%%%%%%

\begin{figure}[t b]
\begin{center}
\begin{minipage}[t]{19cm} % Gruppierung der drei Bilder
\vspace*{-1.6cm}
\begin{minipage}[t]{62mm} % Figure a
\begin{center}
\includegraphics[angle=270,width=60mm]{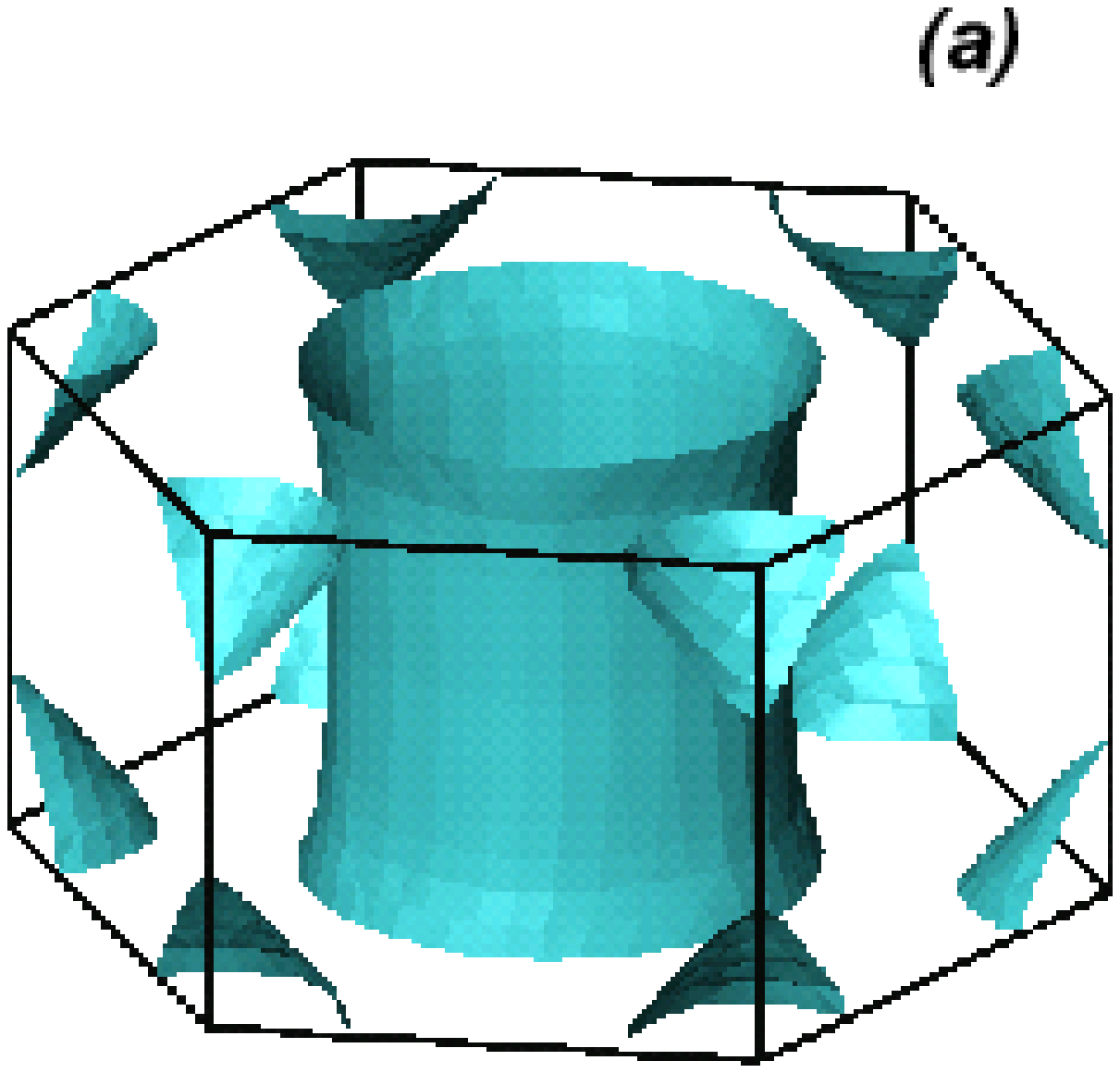}
\end{center}
\end{minipage} % Ende Figure a
\hfill
\begin{minipage}[t]{62mm}  % Figure b
\begin{center}
\includegraphics[angle=270,width=60mm]{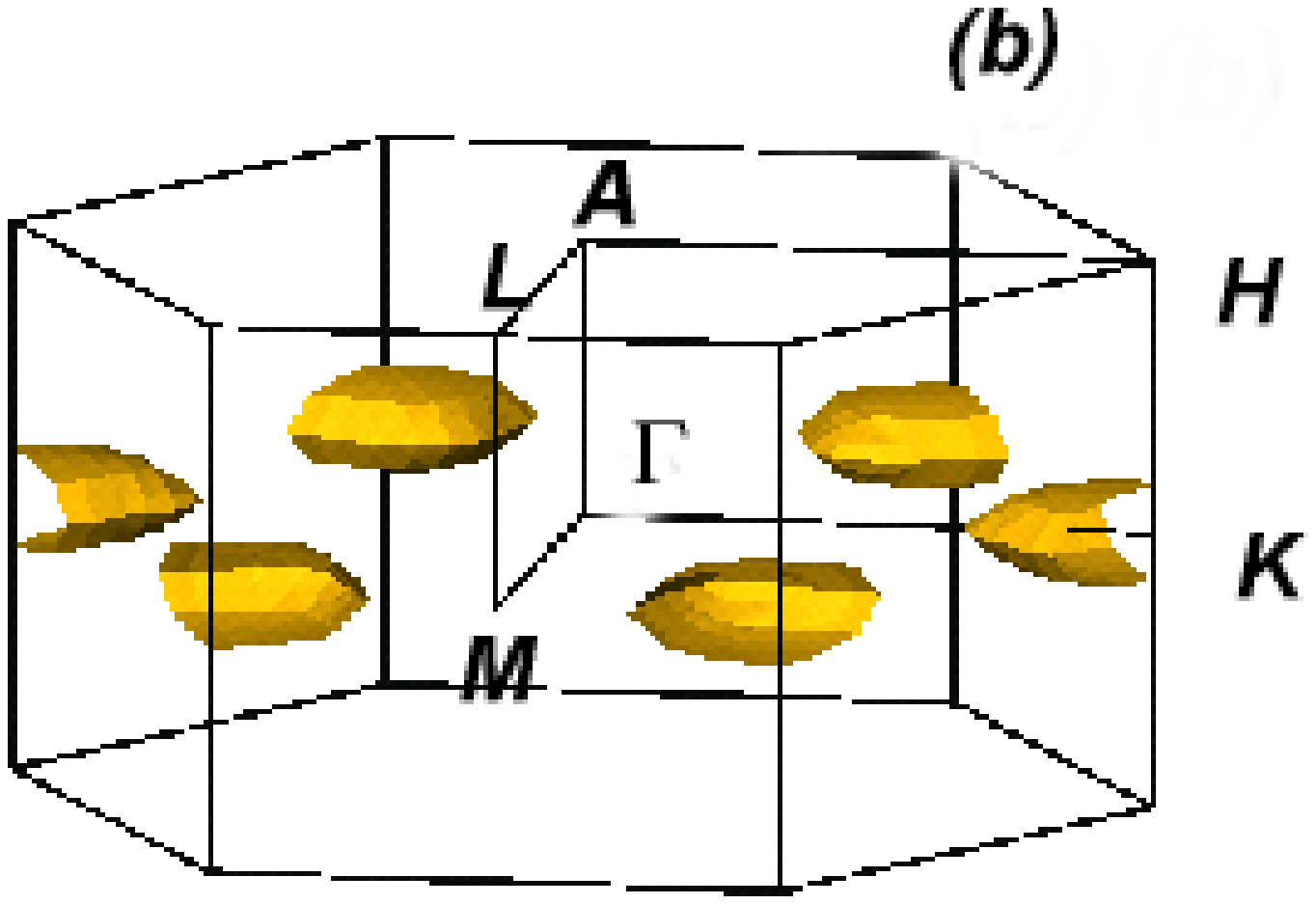}
\end{center}
\end{minipage}  % Ende Figure b
\hfill
\begin{minipage}[t]{62mm}  % Figure c
\begin{center}
\includegraphics[angle=270,width=60mm]{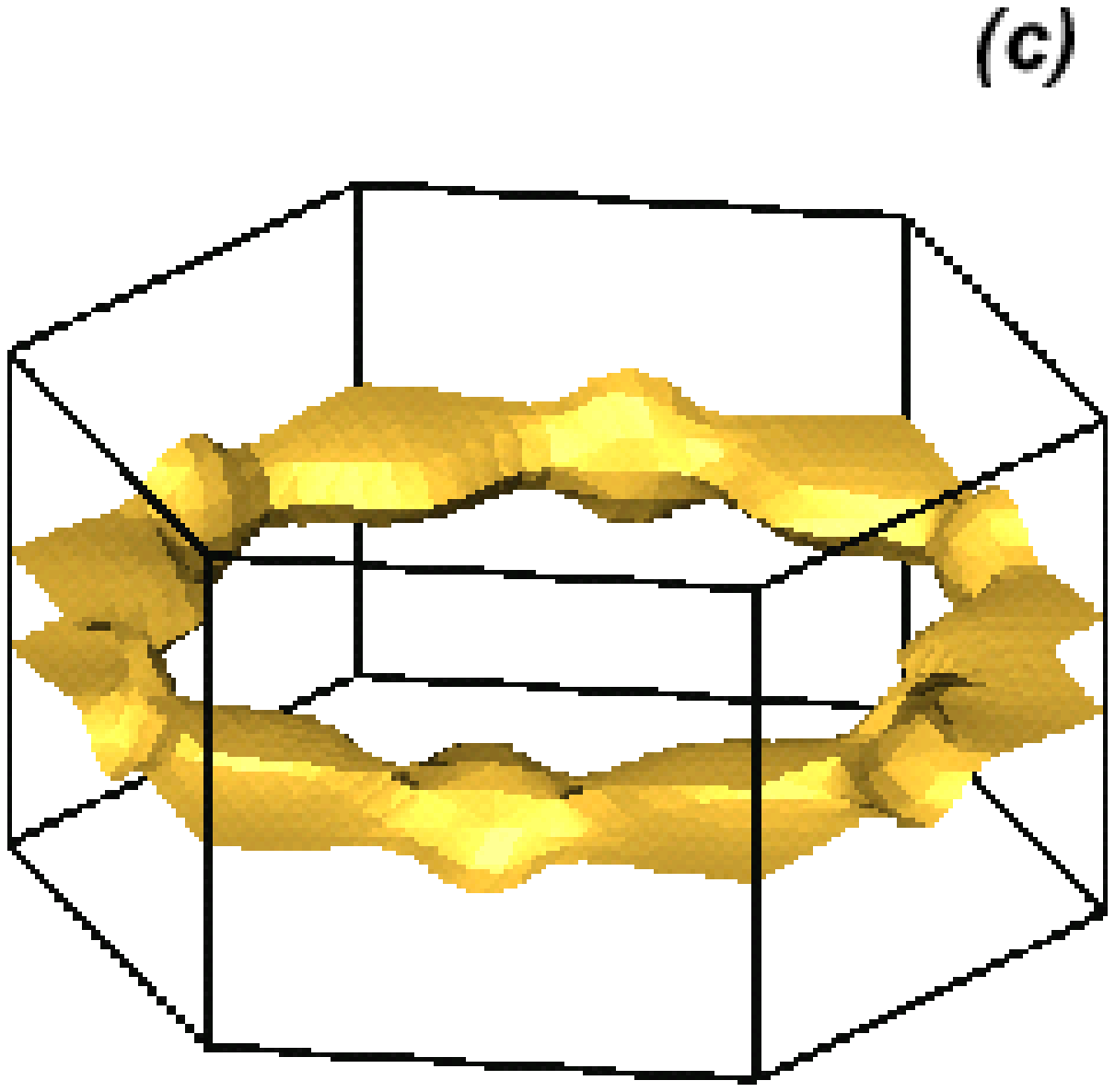}
\end{center}
\end{minipage}  % Ende Figure c
\end{minipage}  % Gruppierung der drei Bilder
\end{center}
\vspace{-1cm}
\caption{Various calculated sheets of the Fermi surface of UPd$_2$Al$_3$.\\
(a):  third sheet (cylinder and
H-centered ellipsoid) (b): second sheet 
(c): second sheet but with a Fermi energy shifted by 40 K.}
\label{fig:Fig2}
\end{figure}

\noindent One notices that the
agreement for the heavy quasiparticle branches is very good. The frequencies
referring to the light parts of the Fermi surface are less well reproduced, but
that is of no surprise. One cannot expect that die LDA reproduces well the
relative shifts of the centers of the light bands which strongly effect the
shape of the corresponding parts of the Fermi surface. The Fermi surface of the
heavy quasiparticles in the antiferromagnetic phase consists of three sheets
two of which are displayed in Figure \ref{fig:Fig2}. 
\end{multicols}
\widetext

%%%%%%%%%%%%%%%%%%%%%%%%%%%%%%%%%%%%%%%%%%%%%%%%%%%%%%%%%%%%%%%%%%%%%%%%%%%%
%Figur 2
%%%%%%%%%%%%%%%%%%%%%%%%%%%%%%%%%%%%%%%%%%%%%%%%%%%%%%%%%%%%%%%%%%%%%%%%%%%%

\begin{multicols}{2}
\narrowtext
\noindent We assign the orbit $\alpha$ to the first sheet
(K-centered ellipsoid) not shown in Fig. \ref{fig:Fig2}. The third
sheet is the most important one and consists of
a corrugated cylinder and a H-centered ellipsoid (see Fig. \ref{fig:Fig2}a) to
which we assign the orbits $\gamma$ and $\beta$ and $\epsilon_2,~ \epsilon_3$,
respectively. The branch $\zeta$ is assigned to the second sheet whose shape
depends sensitively on the position of the Fermi energy (see
Fig. \ref{fig:Fig2}b). For example a shift of the latter by 40 K changes that
sheet from the one shown in Fig. \ref{fig:Fig2}b to that of
Fig. \ref{fig:Fig2}c. But the corresponding changes in the effective mass
remain small. 

In the second step we calculate the multiplet structure of the localized $f^2$
states. This is done within the jj-coupling scheme, because the spin-orbit
splitting is rather large. Thus a 6 x 6 Coulomb matrix has to be diagonalized
for the two-particle states built from $|j=\frac{5}{2},~
j_z=\pm\frac{5}{2}\rangle$  and $|j=\frac{5}{2},~ j_z=\pm\frac{1}{2}\rangle$
for the $f^2$ subshell. The Coulomb matrix elements are calculated following
Condon and Shortley. Inputs are the Slater-Condon
parameters F$^{\rm K}$ (Coulomb integrals) and G$^{\rm K}$ (exchange
integrals). The latter are evaluated with the radial function
R$^U_{f,\frac{5}{2}}$(r) for U as obtained from a self-consistent bandstructure
potential. Thereby the chosen energy is that of the center of gravity of the
5$f$ bands. The required integrations are done within the atomic sphere
surrounding the U ion. We note that we use the same Coulomb matrix as for
UPt$_3$ \cite{ZwYaFu} where also two 5 $f$ electrons with $j_z=\pm\frac{5}{2},~
\pm\frac{1}{2}$ are considered. This is reasonable in view of the fact that in
the nonrelativistic case the matrix elements agree up to two decimals with the
ones computed in \cite{lanthanocenes} for U(C$_8$H$_8$)$_2$, indicating that
they are insensitive to the chemical environment of U. The resulting
eigenstates of the Coulomb matrix are no longer eigenstates of the total
angular momentum J, but remain eigenstates of J$_z$. 

We find a doubly degenerate ground state with J$_z$=$\pm$3. It must be an
eigenstate to J=4 since the Pauli principle requires an even value of J and
J=0,2 are excluded. The states $|j=\frac{5}{2},~
j=\frac{5}{2},~~J=4,~~J_z=\pm3\rangle$ have an overlap of 0.865 with the state
$^3H_4$. The latter follows from Hund's rule when the LS coupling scheme is
applied. The two-fold degeneracy of the ground state is lifted by the
crystalline electric field (CEF) which is acting on the 5$f$ subshell with the
localized electrons. From experiments a $\Gamma_4$ ground state has been
suggested \cite{UPd2Al3CEF} so that the ground- and first excited state are

\begin{equation}
%\label{1}
\mid \Gamma_{3,4} \rangle  =  \frac{1}{\sqrt{2}}~ (\mid J=4; J_z=3 \rangle \pm
\mid J=4;J_z=-3 \rangle)~.
\label{eq:Gamma4}
\end{equation}

For the splitting energy $\tilde{\delta}$ between the two states a value of
20meV has been previously suggested \cite{KFRMGSGLS} while a more recent value
is 7meV \cite{UPd2Al3Nature}. The latter value will be used in the
following. We want to point out that the next higher doublet of the Coulomb
matrix is one with J$_z=\pm$2 with an excitation energy of approximately
0.4 eV. Therefore we may neglect all higher excited states and consider the two
singlets $| \Gamma_4 \rangle$ and $| \Gamma_3 \rangle$ only.

In a third and final step we determine the effective masses which result from
the interaction of the delocalized 5$f$ electrons with the localized one. The
renormalization of the band mass m$_{\rm b}$ is given by 

\begin{equation}
%\label{2}
\frac{m^*}{m_b} = \left. 1 - \frac{\partial \Sigma}{\partial \omega}
\right|_{\omega=0} 
\label{eq:mmb1}
\end{equation}

\noindent where $\Sigma (\omega)$ denotes the local selfenergy of the
delocalized 5$f$ states. The latter is obtained by analytic continuation from
the Matsubara frequencies $\epsilon_{\rm n}=\pi$T(2n+1) at the
temperature T where it is given by

\begin{equation}
%\label{3}
\Sigma(i \epsilon_n) = a^2 M^2 T \sum_{n'} \chi (i \epsilon_n - i
\epsilon_{n'}) g (i \epsilon_{n'}) 
\label{eq:Sigmain}
\end{equation}

\noindent in terms of the local susceptibility

\begin{equation}
%\label{4}
\chi (i \epsilon_n - i \epsilon_{n'}) = - tan h \frac{\tilde{\delta}}{2T}
\frac{2 \tilde{\delta}}{(i \epsilon_n - i \epsilon_{n'})^2 - \tilde{\delta}^2} 
\label{eq:chiepsn}
\end{equation}

\noindent and the local propagator

\begin{equation}
%\label{5}
g(i \epsilon_n)  =  \int dE \frac{N(E)}{i \epsilon_n - E - \Sigma (i
\epsilon_n)}~~~. 
\label{eq:giepslonn}
\end{equation}

\noindent Here 2N(E) is the total density of states at the energy E as obtained
from the bandstructure, when two 5$f$ electrons are kept 
localized. The prefactor $a$ denotes the 5$f$ weight per spin and U atom of the
conduction electron states near E$_{\rm F}$. The matrix element M describes the
transition between the localized states $| \Gamma_4 \rangle$ and $| \Gamma_3
\rangle$ due to the Coulomb interaction U$_{\rm Coul}$ with the delocalized
5$f$ electrons. It is given by

\begin{equation}
%\label{6}
M = \langle f^1; \frac{5}{2}, \frac{3}{2} |\otimes \langle \Gamma_4~
| U_{\rm Coul} |~\Gamma_3 \rangle \otimes |f^1; \frac{5}{2}, \frac{3}{2}
  \rangle 
\label{eq:Mf1Gamma}
\end{equation}

\noindent and is directly obtained from the expection values of the Coulomb
interaction in the 5$f^3$ states. The difference
$\langle f^1;~ \frac{5}{2},~ \frac{3}{2} | \otimes \langle f^2;~ 4,~ 3~ |
U_{\rm Coul} |~ f^2;~ 4,~3 \rangle \otimes |f^1;~ \frac{5}{2},~ \frac{3}{2}
\rangle - \langle f^1;~ \frac{5}{2},~ \frac{3}{2} | \otimes \langle f^2;~ 4,~
-3~ | U_{\rm Coul} |~ f^2;~4,~ -3 \rangle \otimes | f^1;~ \frac{5}{2},~
\frac{3}{2}\rangle$ 

\noindent is -0.38eV. From this we obtain M=-0.19eV. When we want to write M in
form of an exchange coupling between the delocalized and localized 5$f$ states
we need to know the Land\'e factor g. We find g$_{\rm eff}$=0.63 and with this
value an exchange integral of size I$\simeq$1eV. This is of the correct size
for 5$f$ electrons.

In order to evaluate Eq. (\ref{eq:mmb1}) we need to know 
N(E) and $a$. We extract the quantities from LDA calculations with two 5$f$ 
electrons being localized. For simplicity, we model N(E) by a
Lorentzian. The actual value at the Fermi level 
N(0)=2.76 states /(eV cell spin) corresponds to the one found by
\cite{PeTeSzTy}. A value of a$^2$=0.44 is obtained and the value
$\tilde{\delta}$=7meV is used. 

The selfconsistent calculation yields a mass enhancement of 9.6. The resulting
calculated quasiparticle masses are in excellent agreement with experiment (see
Table \ref{tab:effMassH11}).

%%%%%%%%%%%%%%%%%%%%%%%%%%%%%%%%%%%%%%%%%%%%%%%%%%%%%%%%%%%%%%%%%%%%%%%%%%%%
\begin{minipage}[t]{8cm}
\begin{table}
%\vspace{0.5cm}
\caption{Effective masses for H 11 c. The experimental data are taken from
Inada et al. [20].} 

\vspace{0.5cm}
\label{tab:effMassH11}
\begin{tabular}{l c c}
 &  m$^*$ (exp) & m$^*$ (theory)\\[2mm]
\hline
$\zeta$       & 65 & 59.6 \\
$\gamma$      & 33 & 31.9 \\
$\beta$       & 19 & 25.1 \\
$\epsilon_2$  & 18 & 17.4 \\
$\epsilon_3$  & 12 & 13.4 \\
$\alpha$      & 5.7  & 9.6 \\
\end{tabular}
\end{table}
\end{minipage}
%%%%%%%%%%%%%%%%%%%%%%%%%%%%%%%%%%%%%%%%%%%%%%%%%%%%%%%%%%%%%%%%%%%%%%%%%%%%

What remains
to be discussed is the justification for treating two of the 5$f$ electrons in
orbitals $j_z=\frac{5}{2}$ and $\frac{1}{2}$ as localized. As pointed our
before the hybridization of the $j_z=\pm\frac{3}{2}$ orbital with the
neighboring atomic orbitals is larger than the one for the orbitals with
$j_z=\pm\frac{5}{2}$ and $\pm\frac{1}{2}$. So why are we allowed the neglect
the hybridization of the latter orbitals altogether? The answer is found when
the effects of intraatomic correlations are taken into account, i.e., those
resulting from the on-site Coulomb and exchange effects. For a demonstration a
two-site model was treated in Ref. \cite{EfHaRuFuZw} with anisotropic
hybridization between the two U sites. An intermediate valency of 2.5 was used
for the two atoms. Let $| \psi_0 \rangle$ denote the ground state of that
system and $t_\alpha$ the hopping parameters for the different orbitals. It was
found in Ref. \cite{EfHaRuFuZw} that whenever one hopping parameter $t_\alpha$
dominates the others, i.e., $t_\alpha \gg t_{\alpha'}, t_{\alpha''}$, the
corresponding ground state expectation values is

\begin{equation}
%\label{7}
T_{\alpha'} = \frac{\langle \psi_0 \mid t_{\alpha'} c^+_{\alpha'} (1)
c_{\alpha'} (2) \mid \psi_0 \rangle}{\langle \psi_0 \mid \sum\limits_\alpha
t_\alpha c^+_\alpha (1) c_\alpha (2) \mid \psi_0 \rangle} \ll
\frac{t_{\alpha'}}{\sum\limits_\alpha t_\alpha}~~~,
\label{eq:Talpha}
\end{equation}

i.e., the effective hybridization anisotropies are strongly exhanced by
intra-atomic interaction. The operators $c^+_\alpha (i) (c_\alpha (i))$ create
(destroy) an $f$ electron on site $i$ in orbital $j_z=\alpha$. The smaller
hopping matrix elements are therefore suppressed.

%\newpage

%%%%%%%%%%%%%%%%%%%%%%%%%%%%%%%%%%%%%%%%%%%%%%%%%%%%%%%%%%%%%%%%%%%%%%%%%%
% \section*{Acknowledgement} Efremov, Hasselmann, Runge, Yaresko\\
% GZ: Niedersachsen-Israel foundation+ Hospitality of MPI-PKS\\
% We are grateful to S. R. Julian and G. G. Lonzarich for making 
% their experimental data available to us prior to publication.\\
%%%%%%%%%%%%%%%%%%%%%%%%%%%%%%%%%%%%%%%%%%%%%%%%%%%%%%%%%%%%%%%%%%%%%%%%%

\end{multicols}
\end{document}